\documentclass[twocolumn,showpacs,aps,prl,superscriptaddress,letterpaper]{revtex4}

\usepackage{graphicx}
\usepackage{dcolumn}
\usepackage{amsmath}
\usepackage{epsfig}

\long\def\inst#1{\par\nobreak\kern 4pt\nobreak
    {\itshape #1}\par\vskip 10pt plus 3pt minus 3pt}

\RequirePackage{xspace}
\usepackage{relsize}
\def\qqbar {\ensuremath{q\overline q}\xspace}
\def\babar{\mbox{\slshape B\kern-0.1em{\smaller A}\kern-0.1em
    B\kern-0.1em{\smaller A\kern-0.2em R}}}
\def\Bbar    {\kern 0.18em\overline{\kern -0.18em B}{}\xspace}

\def\BB      {\ensuremath{B\Bbar}\xspace} 
\def\Bz      {\ensuremath{B^0}\xspace}
\def\Bzb     {\ensuremath{\Bbar^0}\xspace}
\def\BzBzb   {\ensuremath{\Bz {\kern -0.16em \Bzb}}\xspace}
\def\Bu      {\ensuremath{B^+}\xspace}
\def\Bub     {\ensuremath{B^-}\xspace}

\def\BpBm    {\ensuremath{\Bu {\kern -0.16em \Bub}}\xspace}

\newcommand{\optbar}[1]{\shortstack{{\tiny (\rule[.4ex]{1em}{.1mm})}
  \\ [-.7ex] $#1$}}
\def\BorBbar    {\kern 0.18em\optbar{\kern -0.18em B}{}\xspace}
\def\DorDbar    {\kern 0.18em\optbar{\kern -0.18em D}{}\xspace}
\def\KorKbar    {\kern 0.18em\optbar{\kern -0.18em K}{}\xspace}
\def\CP                {\ensuremath{C\!P}\xspace}
\def\pep2{PEP-II}
\mathchardef\Upsilon="7107
\def\Y#1S{\ensuremath{\Upsilon{(#1S)}}\xspace}

\def\FourS {\Y4S}

\newcommand{\BABARPubYear}     {03}
\newcommand{\BABARPubNumber}  {018}

\newcommand{\SLACPubNumber} {10016}

\begin{document}

\begin{flushleft}
\babar-PUB-\BABARPubYear/\BABARPubNumber\\
SLAC-PUB-\SLACPubNumber
\\[10mm]
\end{flushleft}

\title{
\large \bfseries \boldmath
Rates, Polarizations, and Asymmetries in Charmless Vector-Vector $B$
Meson Decays
}

%
\author{B.~Aubert}
\author{R.~Barate}
\author{D.~Boutigny}
\author{J.-M.~Gaillard}
\author{A.~Hicheur}
\author{Y.~Karyotakis}
\author{J.~P.~Lees}
\author{P.~Robbe}
\author{V.~Tisserand}
\author{A.~Zghiche}
\affiliation{Laboratoire de Physique des Particules, F-74941 Annecy-le-Vieux, France }
% ...
\collaboration{The \babar\ Collaboration}
\noaffiliation

\date{July 10, 2003}


\begin{abstract}
With a sample of approximately 89 million $\BB$
pairs collected with the $\babar$ detector, 
we perform a search for
$B$ meson decays 
into pairs of charmless vector mesons 
($\phi$, $\rho$, and $K^*$).
We measure the branching 
fractions, determine the degree of longitudinal 
polarization, and search for $\CP$ violation asymmetries 
in the processes $B^+\to\phi K^{*+}$, $B^0\to\phi K^{*0}$, 
$B^+\to\rho^0 K^{*+}$, and $B^+\to\rho^0 \rho^+$.
We also set an upper limit on the branching fraction
for the decay $B^0\to\rho^0\rho^0$.
\end{abstract}


\vspace{-0.5cm}

\maketitle


Charmless $B$ meson decays provide an opportunity to measure 
the weak-interaction phases arising from the elements of the 
Cabibbo-Kobayashi-Maskawa (CKM) quark-mixing matrix~\cite{Kobayashi} 
and to search for phenomena outside the standard model,  
including charged Higgs bosons and supersymmetric 
particles~\cite{newphys}.

The decays to two vector particles are of special interest 
because their angular distributions reflect both
strong and weak interaction dynamics~\cite{bvv}. 
The asymmetries constructed from the number of $B$ decays
with each flavor and with each sign of a triple product 
are sensitive to $\CP$ violation or to final state 
interactions (FSI)~\cite{tpc}. The triple product is defined as 
$(\mathbf{q_1}-\mathbf{q_2})\cdot\mathbf{p_1}\times\mathbf{p_2}$,
where {$\mathbf{q_1}$} and {$\mathbf{q_2}$} are the momenta
of the two vector particles in the $B$ frame and 
{$\mathbf{p_1}$} and {$\mathbf{p_2}$} represent 
their polarization vectors.

The first evidence for the decays of $B$ mesons to pairs 
of charmless vector mesons 
was provided by the CLEO~\cite{cleo} and \babar~\cite{prlphik}
experiments with the observation of $B\to{\phi K^{*}}$ decays.
The CLEO experiment also set upper limits on the $B$ decay rates
for several other vector-vector final states \cite{cleorhokst}.
The BELLE experiment recently announced observation of
$B^+\to\rho^0\rho^+$ \cite{belle}.


In this analysis, we use
the data collected with the \babar\ detector~\cite{babar}
at the \pep2 asymmetric-energy $e^+e^-$ collider~\cite{pep}.
These data represent an integrated luminosity 
of 81.9~fb$^{-1}$, corresponding to 88.9 million 
$\BB$ pairs, at the $\FourS$ resonance
(on-resonance) and 9.6~fb$^{-1}$ approximately 
40~MeV below this energy (off-resonance). 
The $\FourS$ resonance occurs at the $e^+e^-$ 
center-of-mass (c.m.) energy, $\sqrt{s}$, of~10.58~GeV.

Charged-particle momenta are measured in a tracking system 
that is a combination of a silicon vertex tracker (SVT) consisting 
of five double-sided detectors and a 40-layer central drift chamber 
(DCH), both operating in a 1.5 T solenoidal magnetic field. 
Charged-particle identification is provided by the 
energy loss ($dE/dx$) in the tracking devices (SVT and DCH) and
by an internally reflecting ring-imaging Cherenkov detector 
(DIRC) covering the central region. 
Photons are detected by a CsI(Tl) electromagnetic calorimeter.


We search for charmless vector-vector $B$ meson decays 
involving $\phi$, $\rho$, and $K^*(892)$ resonances.
The event selection and analysis technique have been 
discussed earlier~\cite{prlphik}. 
We fully reconstruct
the charged and neutral decay products including
the intermediate states $\phi\to K^+K^-$, 
$K^{*0}\to K^+\pi^-$ and $K^0\pi^0$,
$K^{*+}\to K^+\pi^0$ and $K^0\pi^+$, 
$\rho^0\to \pi^+\pi^-$, $\rho^+\to \pi^+\pi^0$, with
$\pi^0\rightarrow \gamma\gamma$ and
$K^0\rightarrow K^0_S\rightarrow\pi^+\pi^-$,
where inclusion of the charge conjugate states is implied.
Candidate charged tracks are required to originate 
from the interaction point.
Looser criteria are applied to tracks forming 
$K^0_S$ candidates, which are required to satisfy 
$|m_{\pi^+\pi^-} - m_{K^0}|<$ 12~MeV  
with the cosine of the angle between their reconstructed flight and
momentum directions greater than 0.995, and the measured proper 
decay time greater than 5 times its uncertainty.
Charged-particle identification provides separation
of kaon tracks from pion and proton tracks.

We reconstruct $\pi^0$ mesons from pairs of photons, each with 
a minimum energy of 30 MeV. 
The invariant mass of the $\pi^0$ candidates is required 
to be within 15 MeV of the nominal mass.
The helicity angle of a $\phi$, $K^*$, or $\rho$
is defined as the angle between the momentum 
({$\mathbf{p_1}$} or {$\mathbf{p_2}$})
of one of its two daughters 
($K^+$, $K$, or $\pi^+$, respectively) 
in the resonance rest frame
and the momentum ({$\mathbf{q_1}$} or {$\mathbf{q_2}$})
of the resonance in the $B$ frame.
To suppress combinatorial background with low-energy 
$\pi^0$ candidates we restrict the $K^{*}\rightarrow K\pi^0$ 
and $\rho^{+}\rightarrow \pi^+\pi^0$ helicity angle $\theta$
range to $\cos\theta < +0.5$. 

We identify $B$ meson candidates kinematically
using two nearly independent variables \cite{babar}:
the beam-energy-substituted mass $m_{\rm{ES}} =$ 
$[{ (s/2 + \mathbf{p}_i \cdot \mathbf{p}_B)^2 / E_i^2 - 
\mathbf{p}_B^{\,2} }]^{1/2}$ and the energy difference
$\Delta{E}=(E_iE_B-\mathbf{p}_i$$\cdot$$\mathbf{p}_B-s/2)/\sqrt{s}$,
where $(E_i,\mathbf{p}_i)$ is the initial state four-momentum
obtained from the beam momenta, and $(E_B,\mathbf{p}_B)$
is the four-momentum of the reconstructed $B$ candidate.
Our initial selection requires $m_{\rm{ES}}>5.2$ GeV
and $|\Delta E|<0.2$~GeV.

To reject the dominant quark-antiquark continuum background,
we require $|\cos\theta_T| < 0.8$, where $\theta_T$ 
is the angle between the $B$-candidate thrust axis
and that of the rest of the tracks and neutral clusters in
the event, calculated in the c.m. frame. 
We also construct a Fisher discriminant that combines eleven 
event-shape variables defined in the c.m. 
frame~\cite{prlphik, CLEO-fisher}.

Monte Carlo (MC) simulation \cite{geant} demonstrates that 
contamination from other $B$ decays is negligible for the 
modes with a narrow $\phi$ resonance and is relatively small 
for other charmless $B$ decay modes. We achieve further 
suppression of $B$-decay background by removing signal
candidates that have decay products consistent with
$D\rightarrow K\pi, K\pi\pi$ decays. The remaining small 
background coming from $B$ decays (about 6\% of the total 
background) is taken into account in the fit described below. 
In this analysis we do not explicitly provide a fit component
for other partial waves with the same final-state particles
selected within vector resonance mass windows.


We use an unbinned, extended maximum-likelihood (ML) 
fit~\cite{prlphik} to extract
signal yields, asymmetries, and angular polarizations simultaneously. 
We define the likelihood ${\cal L}_i$ for each event candidate $i$
as the sum of $n_{jk}{\cal P}_{j}(\vec{x}_{i};\vec{\alpha})$
over three event categories $j$,
where ${\cal P}_{j}(\vec{x}_{i};\vec{\alpha})$ are the probability 
density functions (p.d.f.) for measured variables $\vec{x}_{i}$,
$n_{jk}$ are the yields to be extracted from the fit,
and $k$ is the measured tag (1 or 2, as defined 
for asymmetry measurements later).
There are three categories: signal ($j=1$), continuum~$\qqbar$
($j=2$), and $\BB$ combinatorial background ($j=3$).
The fixed numbers $\vec{\alpha}$ parameterize the expected 
distributions of measured variables in each category.
They are extracted from MC simulation, on-resonance
$\Delta E$-$m_{\rm{ES}}$ sidebands, and off-resonance
data. 

\begingroup
\begin{table*}[btp]
\caption
{Summary of results for the measured $B$-decay modes;
$\varepsilon$ denotes the reconstruction efficiency and
$\varepsilon_{\rm{tot}}$ 
the total efficiency including daughter branching fractions,
$n_{\rm sig}$ is the fitted number of signal events, 
${\cal B}$ is the branching fraction, 
$f_L$ is the longitudinal polarization, and
${\cal A}_{\CP}$ is the signal charge asymmetry.
The decay channels of $K^*$ are shown when more than one
final state is measured for the same $B$ decay mode.
All results include systematic errors, which are quoted 
following the statistical errors. The errors are combined
for the reconstruction efficiency. 
The upper limit on the $B^0\to\rho^0\rho^0$ branching fraction
is given at $90\%$ confidence level including systematic
uncertainties and conservatively assuming the efficiency 
for $f_L=1$.}
\label{tab:results}
\footnotesize
\begin{center}
\begin{ruledtabular}
\setlength{\extrarowheight}{1.5pt}
\begin{tabular}{lcccccc}
\vspace{-2.5mm}&&&\\
Mode & $\varepsilon$ (\%) & $\varepsilon_{\rm tot}$ (\%)
 & $n_{\rm sig}$ 
 & ${\cal B}$ ($\times 10^{-6}$) & $f_L$ & ${\cal A}_{\CP}$ \cr
\vspace{-2.5mm}&&&\\
\hline
\vspace{-2.5mm}&&&\\
$\phi K^{*+}$ 
 & -- & 5.0 & -- & $12.7^{~+2.2}_{~-2.0}\pm 1.1$  
 & $0.46\pm 0.12\pm 0.03$ & $+0.16\pm 0.17\pm 0.03$ \cr
\vspace{-2.5mm}&&&\\
~~~$\rightarrow$${K^0\pi^+}$ 
 & $23.9\pm 2.1$ & 2.7 & $33.3^{+7.2}_{-6.4}\pm 1.2$  & $13.9^{~+3.0}_{~-2.7}\pm 1.2$  
 & $0.50^{~+0.14}_{~-0.15}\pm 0.03$ & $-0.02\pm 0.20\pm 0.03$ \cr
\vspace{-2.5mm}&&&\\
~~~$\rightarrow$${K^+\pi^0}$ 
 & $14.3\pm 1.4$ & 2.3 & $22.3^{+7.5}_{-6.5}\pm 3.2$  & $10.7^{~+3.6}_{~-3.1}\pm 1.8$  
 & $0.40^{~+0.20}_{~-0.19}\pm 0.06$ & $+0.63^{~+0.25}_{~-0.31}\pm 0.05$  \cr
\vspace{-2.5mm}&&&\\
\hline
\vspace{-2.5mm}&&&\\
$\phi K^{*0}$ 
 & -- & 10.3 & -- & $11.2\pm 1.3\pm 0.8$  
 & $0.65\pm 0.07\pm 0.02$ & $+0.04\pm 0.12\pm 0.02$  \cr
\vspace{-2.5mm}&&&\\
~~~$\rightarrow$${K^+\pi^-}$ 
 & $29.7\pm 2.6$ & 9.7 & $101^{~+12}_{~-11}\pm 3$  & $11.7\pm 1.4\pm 0.8$  
 & $0.64\pm 0.07\pm 0.02$ & $+0.04\pm 0.12\pm 0.02$  \cr
\vspace{-2.5mm}&&&\\
~~~$\rightarrow$${K^0\pi^0}$ 
 & $10.5\pm 1.0$ & 0.6 & $2.0^{+3.4}_{-1.3}\pm 0.6$ & $~3.8^{~+6.6}_{~-2.5}\pm 1.1$  
 & $1.00^{~+0.00}_{~-0.66}\pm 0.25$ & --  \cr
\vspace{-2.5mm}&&&\\
\hline
\vspace{-2.5mm}&&&\\
$\rho^0 K^{*+}$ 
 & -- & 4.8 & -- & $10.6^{+3.0}_{-2.6}\pm 2.4$  
 & $0.96^{~+0.04}_{~-0.15}\pm 0.04$ & $+0.20^{~+0.32}_{~-0.29}\pm 0.04$  \cr  
\vspace{-2.5mm}&&&\\
~~~$\rightarrow$${K^0\pi^+}$ 
 & $12.3\pm 2.0$ & 2.8 & $35.7^{+11.8}_{-11.0}\pm 3.6$  & $14.3^{+4.7}_{-4.4}\pm 2.9$ 
 & $0.90^{~+0.10}_{~-0.16}\pm 0.04$ & $+0.17^{~+0.34}_{~-0.31}\pm 0.04$  \cr
\vspace{-2.5mm}&&&\\
~~~$\rightarrow$${K^+\pi^0}$ 
 & $6.0\pm 1.4$ & 2.0 & $8.5^{+8.2}_{-6.6}\pm 5.2$  & $4.8^{+4.6}_{-3.7}\pm 3.2$ 
 & $1.00^{~+0.00}_{~-0.20}\pm 0.03$ & $+0.28^{~+0.72}_{~-0.82}\pm 0.19$  \cr
\vspace{-2.5mm}&&&\\
\hline
\vspace{-2.5mm}&&&\\
$\rho^0\rho^+$ 
 & $4.7\pm 0.9$ & 4.6 & $93^{~+24}_{~-22}\pm 10$  & $22.5^{+5.7}_{-5.4}\pm 5.8$ 
 & $0.97^{~+0.03}_{~-0.07}\pm 0.04$ & $-0.19\pm 0.23\pm 0.03$  \cr
\vspace{-2.5mm}&&&\\
\hline
\vspace{-2.5mm}&&&\\
$\rho^0\rho^0$ 
 & $17.6\pm 1.5$ & 17.6 & $9.7^{+11.9}_{-9.4}\pm 2.0$  & $< 2.1$ & -- & -- \cr
\vspace{-2.5mm}&&&\\
\end{tabular}
\end{ruledtabular}
\end{center}
\end{table*}
\endgroup

The fit input variables $\vec{x}_{i}$ are $\Delta E$, 
$m_{\rm{ES}}$, Fisher discriminant, invariant masses 
of the candidate $K^*$ and $\phi$ (or $\rho$) resonances, 
and the $K^*$ and $\phi$ (or $\rho$) helicity angles
$\theta_{\rm 1}$ and $\theta_{\rm 2}$.
The correlations among the fit input variables 
in the data and signal MC are found to be small 
(typically less than $5\%$), except for angular 
correlations in the signal as discussed below.
The p.d.f. ${\cal P}_{j}(\vec{x}_{i};\vec{\alpha})$ 
for a given candidate $i$ is the product of the p.d.f.'s
for each of the variables and a joint p.d.f. for the  
helicity angles, which accounts for the angular correlations 
in the signal and for detector acceptance effects.
We integrate over the angle between the decay planes 
of the two vector-particle decays, leaving a p.d.f. that depends 
only on the two helicity angles and the unknown longitudinal 
polarization fraction $f_L\equiv{\Gamma_L}/{\Gamma}$.
The differential decay width~\cite{bvv}
${d^2\Gamma / d\cos \theta_1 d\cos \theta_2}$ is 
\begin{eqnarray}
{9\Gamma \over 4} \left \{ {1 \over 4} (1 - f_L)
\sin^2 \theta_1 \sin^2 \theta_2 + f_L \cos^2 \theta_1 \cos^2 \theta_2 \right\}.
\label{eq:helicityintegr}
\end{eqnarray}

To describe the signal distributions, we use Gaussian functions 
for the parameterization of the p.d.f.'s for $\Delta E$ and
$m_{\rm{ES}}$, and a relativistic $P$-wave Breit-Wigner distribution,
convoluted with a Gaussian resolution function, for the resonance masses.
For the background, we use low-degree polynomials or, in the case 
of $m_{\rm{ES}}$, an empirical phase-space function~\cite{argus}.
The background parameterizations for resonance masses also include 
a resonant component to account for resonance production in the 
continuum. The background helicity-angle distribution is 
again separated into contributions from combinatorial background
and from real vector mesons, both described by polynomials.
The p.d.f. for the Fisher 
discriminant is represented by a Gaussian distribution 
with different widths above and below the mean.

We denote $Q_{\rm tp}$ as the sign of the triple product
and $Q_{\rm ch}$ as the $B$-flavor sign ($Q_{\rm ch}=+1$ 
for $\Bbar$ and $Q_{\rm ch}=-1$ for $B$).
The charged $B$ is intrinsically flavor-tagged. 
The flavor of a neutral $B$ is determined from 
the charge of the kaon in the final states with
the $K^{*0}\to K^+\pi^-$, but is undetermined for 
the decay mode $K^{*0}\to K^0\pi^0$ and for the 
decay $B^0\to\rho^0\rho^0$.

We rewrite the event yields $n_{jk}$ ($k$=1,2)
in each category $j$ in terms of 
the asymmetry ${\cal A}_j$ and the total event yield $n_{j}$:
$n_{j1} = n_{j}\times(1 + {\cal A}_j)/2$ and
$n_{j2} = n_{j}\times(1 - {\cal A}_j)/2$.
We define three signal asymmetries using the tags~$k$: 
${\cal A}_{\CP}$ 
($k = 1$ for $Q_{\rm ch}>0$, $k = 2$ for $Q_{\rm ch}<0$),
${\cal A}_{\rm tp}$  
($k = 1$ for $Q_{\rm ch}\times Q_{\rm tp}>0$,
 $k = 2$ for $Q_{\rm ch}\times Q_{\rm tp}<0$),
and ${\cal A}_{\rm sp}$ 
($k = 1$ for $Q_{\rm tp}>0$, $k = 2$ for $Q_{\rm tp}<0$).
A non-zero value for ${\cal A}_{\CP}$ would provide 
evidence for direct-$\CP$ violation, 
non-zero ${\cal A}_{\rm tp}$ indicates $\CP$ violation
even in the absence of FSI, and ${\cal A}_{\rm sp}$ is 
sensitive to strong-interaction phases~\cite{tpc}.

We allow for multiple candidates in a given event by assigning 
to each a weight of $1/N_i$, where $N_i$ is the number of 
candidates in the same event.
The extended likelihood for a sample 
of $N_{\rm cand}$ candidates is 
\begin{equation}
{\cal L} = \exp\left(-\sum_{j=1}^{3} n_{j}\right)\, 
\prod_{i=1}^{N_{\rm cand}} 
\exp\left(\frac{\ln{\cal L}_i}{N_i}\right) .
\label{eq:likel}
\end{equation}

The event yields $n_j$, asymmetries ${\cal A}_j$, and
polarization $f_L$ are obtained by minimizing 
the quantity $\chi^2\equiv -2\ln{\cal L}$. 
The dependence of $\chi^2$ on a fit parameter
$n_j$, ${\cal A}_j$, or $f_L$ is obtained with the other
fit parameters floating, their values are constrained 
to the physical range $0\le f_L\le 1$ and $0\le n_j$.
We quote statistical errors corresponding to unit
change in $\chi^2$.
When more than one $K^*$ decay channel is measured for the 
same $B$ decay, the channels are combined by adding their 
$\chi^2$ distributions for $n_j$, ${\cal A}_j$, or $f_L$.
The statistical significance of a signal is defined as the 
square root of the change in $\chi^2$ when constraining 
the number of signal events to zero in the likelihood fit.
If no significant event yield is observed, we quote an upper 
limit for the branching fraction obtained by integrating the 
normalized likelihood distribution.
Performance of the ML fit is tested 
with generated MC and control samples.

\begin{figure}[hbt]
\setlength{\epsfxsize}{1.0\linewidth}\leavevmode\epsfbox{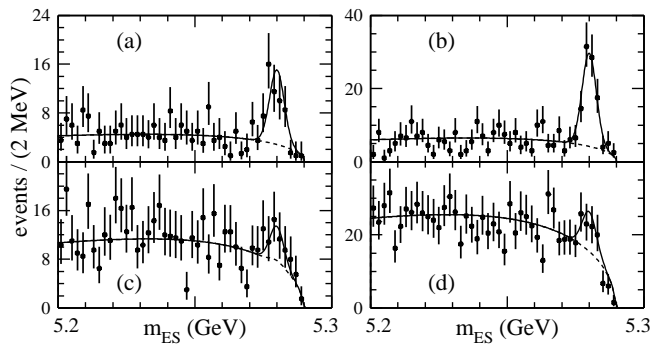}
\caption{\label{fig:mbproj_phiKst} 
Projections of the multidimensional fit
onto the variable $m_{\rm{ES}}$ for 
(a) $B^+\rightarrow\phi K^{*+}$, 
(b) $B^0\rightarrow\phi K^{*0}$,
(c) $B^+\rightarrow\rho^0 K^{*+}$, and 
(d) $B^+\rightarrow\rho^0 \rho^{+}$ candidates
after a requirement on 
the signal-to-background probability ratio
${\cal P}_{\rm{sig}}/{\cal P}_{\rm{bkg}}$ with the 
p.d.f. for $m_{\rm{ES}}$ excluded.
The points with error bars show the data,
the solid (dashed) line shows the signal-plus-background 
(background only) p.d.f. projection.
}
\end{figure}
\begin{figure}[hbt]
\setlength{\epsfxsize}{1.0\linewidth}\leavevmode\epsfbox{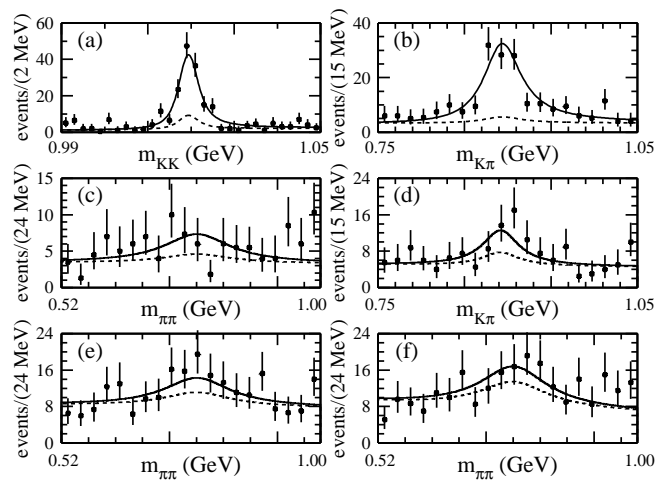}
\caption{\label{fig:massproj} 
Invariant mass projections  
(a) $\phi$, (b) $K^{*}$ for $B\rightarrow\phi K^{*}$;
(c) $\rho^0$, (d) $K^{*+}$ for $B^+\rightarrow\rho^0 K^{*+}$;
(e) $\rho^0$, (f) $\rho^{+}$ for $B^+\rightarrow\rho^0 \rho^{+}$
candidates after a requirement on 
the signal-to-background probability ratio
${\cal P}_{\rm{sig}}/{\cal P}_{\rm{bkg}}$ with the 
p.d.f. for mass excluded.
For point and line definitions see Fig.~\ref{fig:mbproj_phiKst}.
}
\end{figure}

The results of our maximum likelihood fits are summarized in 
Table~\ref{tab:results}. For the branching fractions, 
we assume equal production rates of $\BzBzb$ and $\BpBm$.
We find significant signals in $\rho^0 K^{*+}$ (4.8$\sigma$),
$\rho^0\rho^+$ (6.1$\sigma$), and in both $\phi K^*$ 
(above 10$\sigma$ each) decay modes.
We measure the charge asymmetries and longitudinal 
polarizations in the above modes. The projections of 
the fit results are shown 
in Fig.~\ref{fig:mbproj_phiKst} and~\ref{fig:massproj}.
The asymmetries involving triple products are obtained 
from separate fits.
The results are shown in Table~\ref{tab:results2}.


Systematic uncertainties in the ML fit originate from assumptions 
about the p.d.f.'s.
We vary the p.d.f. parameters within their respective uncertainties,
and derive the associated systematic errors.
The signals remain statistically significant under
these variations. Additional systematic errors in the number
of signal events originate from uncertainty in the 
background component for $\rho K^*$ that peaks in 
$m_{\rm ES}$, where we take the uncertainties to be 
the estimated values.

\begingroup
\begin{table}[b]
\caption
{Summary of asymmetry results with triple products discussed
in the text.}
\label{tab:results2}
\footnotesize
\begin{center}
\begin{ruledtabular}
\setlength{\extrarowheight}{1.5pt}
\begin{tabular}{lcc}
\vspace{-2.5mm}&&\\
Mode & ${\cal A}_{\rm tp}$ & ${\cal A}_{\rm sp}$ \cr
\vspace{-2.5mm}&&\\
\hline
\vspace{-2.5mm}&&\\
$\phi K^{*+}$ 
 & $-0.02\pm 0.18\pm 0.03$ & $-0.04\pm 0.18\pm 0.03$ \cr
\vspace{-2.5mm}&&\\
$\phi K^{*0}$
 & $+0.06\pm 0.12\pm 0.02$ & $+0.07\pm 0.12\pm 0.02$  \cr
\vspace{-2.5mm}&&\\
$\rho^0 K^{*+}$
 & $+0.03\pm 0.29\pm 0.03$ & $+0.28^{~+0.38}_{~-0.33}\pm 0.04$\cr
\vspace{-2.5mm}&&\\
$\rho^0\rho^+$
 & $+0.09\pm 0.24\pm 0.04$ & $-0.23\pm 0.24\pm 0.04$ \cr
\vspace{-2.5mm}&&\\
\end{tabular}
\end{ruledtabular}
\end{center}
\end{table}
\endgroup

The systematic errors in the efficiencies are for track 
finding (0.8\% per track), particle identification (2\% per track), 
and $K^0_S$ and $\pi^0$ reconstruction (5\% each). 
Other minor systematic effects are from event-selection criteria, 
daughter branching fractions~\cite{pdg}, MC statistics, 
and number of $B$ mesons.
We account for the fake combinations in signal events passing 
the selection criteria with a systematic uncertainty of
3--12\%, depending on the mode. 
The reconstruction efficiency depends on the decay 
polarization. We calculate the efficiencies using 
the polarization measured in each decay mode
\footnote{Preliminary $\babar$ results prior to polarization
measurements assumed $f_L=0.5$, leading to a smaller branching
fraction. The systematic error on the branching ratio due 
to the unknown polarization was underestimated
in the preliminary result.}
(combined for the two $\phi K^{*}$ modes)
and assign a systematic error corresponding to the total 
polarization measurement error.
The $B^0\to\rho^0\rho^0$ branching fraction limit 
incorporates uncertainties in the p.d.f.'s and in the 
reconstruction efficiency, 
while we conservatively assume $f_L=1$ for the efficiency
(which is 29\% for $f_L=0$ and 18\% for $f_L=1$).

In the polarization and asymmetry measurements, 
we again include systematic errors from p.d.f. variations 
that account for uncertainties in the detector acceptance and 
background parameterizations.
The biases from the finite resolution in helicity angle measurement 
and dilution due to the presence of the fake combinations are
studied with MC simulation and are accounted for with a 
systematic error of 0.02 for polarization.
We find the uncertainty on the charge asymmetry due to the 
track reconstruction efficiency to be less than 0.02~\cite{prlphik}.
The asymmetry measurements are corrected by the small dilution 
factors.


In summary, we have observed the decays 
$B^+\to\phi K^{*+}$, $B^0\to\phi K^{*0}$,
$B^+\to\rho^0 K^{*+}$, and $B^+\to\rho^0\rho^{+}$,
measured their branching fractions and longitudinal 
polarizations, and looked for asymmetries
sensitive to $\CP$ violation and FSI. 
These results supersede the earlier $\babar$ measurements 
of the $B\to\phi K^{*}$~\cite{prlphik}.
Our asymmetry results rule out a significant part of the 
physical region, providing constraints on models with 
hypothetical particles, 
but are not yet sufficiently precise to 
allow detailed comparison with standard model predictions.
Our measurement of longitudinal polarization is of 
interest for the study of decay dynamics.


We are grateful for the excellent luminosity and machine conditions
provided by our \pep2\ colleagues, 
and for the substantial dedicated effort from
the computing organizations that support \babar.
The collaborating institutions wish to thank 
SLAC for its support and kind hospitality. 
This work is supported by
DOE
and NSF (USA),
NSERC (Canada),
IHEP (China),
CEA and
CNRS-IN2P3
(France),
BMBF and DFG
(Germany),
INFN (Italy),
FOM (The Netherlands),
NFR (Norway),
MIST (Russia), and
PPARC (United Kingdom). 
Individuals have received support from the 
A.~P.~Sloan Foundation, 
Research Corporation,
and Alexander von Humboldt Foundation.


\bibliographystyle{h-physrev2-original}   %

\begin{thebibliography}{99}

\bibitem{Kobayashi}
\label{ref:Kobayashi}
M.~Kobayashi, T. Maskawa, Prog. Theor. Phys. {\bf 49}, 652 (1973).

\bibitem{newphys}
\label{ref:newphys}
Y. Grossman, M.P. Worah,
Phys.\ Lett.\ B {\bf 395}, 241 (1997);
D. London, A. Soni,
Phys.\ Lett.\ B {\bf 407}, 61 (1997).

\bibitem{bvv}
G. Kramer, W.F. Palmer, Phys.\ Rev.\ D {\bf 45}, 193 (1992);
H.-Y.~Cheng, K.-C.~Yang,
Phys.\ Lett.\ B {\bf 511}, 40 (2001);
C.-H. Chen, Y.-Y. Keum, H.-n. Li,
Phys.\ Rev.\ D {\bf 66}, 054013 (2002).

\bibitem{tpc}
G.~Valencia, Phys.\ Rev.\ D {\bf 39}, 3339 (1989);
W.~Bensalem, D.~London, Phys.\ Rev.\ D {\bf 64}, 116003 (2001).

\bibitem{cleo}
\label{ref:cleo}
CLEO Collaboration, R.A.~Briere {\it et al.}, 
Phys.\ Rev.\ Lett. {\bf 86}, 3718 (2001).

\bibitem{prlphik}
$\babar$ Collaboration, B.~Aubert {\it et al.},
Phys.\ Rev.\ Lett. {\bf 87}, 151801 (2001);
$\babar$ Collaboration, B.~Aubert {\it et al.},
Phys.\ Rev.\ D {\bf 65}, 051101 (2002).

\bibitem{cleorhokst}
CLEO Collaboration, R. Godang {\it et al.},
Phys.\ Rev.\ Lett. {\bf 88}, 021802 (2002).

\bibitem{belle}
BELLE Collaboration, J. Zhang {\it et al.},
BELLE-2003-6, hep-ex/0306007,
submitted to Phys.\ Rev.\ Lett.

\bibitem{babar}
\babar\ Collaboration, B.~Aubert {\it et al.},
Nucl.\ Instrum.\ Methods {\bf A479}, 1 (2002).

\bibitem{pep} 
PEP-II Conceptual Design Report, SLAC-R-418 (1993).

\bibitem{CLEO-fisher}
CLEO Collaboration,
D.M.~Asner {\it et al.}, 
Phys.\ Rev.\ D {\bf 53}, 1039 (1996).

\bibitem{geant}
The \babar\ detector Monte Carlo 
simulation is based on GEANT:
R.~Brun {\it et al.}, CERN DD/EE/84-1.

\bibitem{argus}
ARGUS Collaboration, H.~Albrecht {\it et al.}, Phys.\ Lett.\ B {\bf 241}, 278 (1990).

\bibitem{pdg} 
Particle Data Group,
K. Hagiwara {\it et al.}, Phys.\ Rev.\ D {\bf 66}, 010001 (2002).

\end{thebibliography}

\end{document}